\documentclass[noshowpacs,amsmath,twocolumn,
aps,prb]
{revtex4-1}
\usepackage{xcolor}
\usepackage{amsmath}
\usepackage{graphicx}
\usepackage{dcolumn}
\usepackage{bm}
\usepackage{booktabs}

\usepackage{color}
\DeclareGraphicsExtensions{.pdf,.eps,.png,.jpg,.mps} 
\usepackage{hyperref}
\hypersetup{
 colorlinks=true,
 linkcolor=blue,
 citecolor=blue,
 filecolor=blue, 
 urlcolor=blue,
}
\begin{document}

\title{Coherence memory and amnesia in a mode-locked laser}


\author{Bo Cao}
\affiliation{State Key Laboratory of Precision Measurement Technology and Instruments, Department of Precision Instruments, Tsinghua University, Beijing 100084, China}

\author{Zhongshu Liu}%
\affiliation{State Key Laboratory of Precision Measurement Technology and Instruments, Department of Precision Instruments, Tsinghua University, Beijing 100084, China}

\author{Chenxin Gao}%
\affiliation{State Key Laboratory of Precision Measurement Technology and Instruments, Department of Precision Instruments, Tsinghua University, Beijing 100084, China}

\author{Changxi Yang}%
\affiliation{State Key Laboratory of Precision Measurement Technology and Instruments, Department of Precision Instruments, Tsinghua University, Beijing 100084, China}

\author{Chengying Bao}
\email{cbao@tsinghua.edu.cn}
\affiliation{State Key Laboratory of Precision Measurement Technology and Instruments, Department of Precision Instruments, Tsinghua University, Beijing 100084, China}

\date{\today}

\begin{abstract}

\end{abstract}

\maketitle



{\bf Self-organization of temporal modes in mode-locked lasers usually starts from quantum noise. In this process, incoherent spontaneous emission is steered into coherent ultrashort pulses by dissipation and nonlinearity. In this work, we investigated self-organization dynamics in a mode-locked Mamyshev oscillator starting from coherent pulse seeds as opposed to quantum noise. We observed that the coherence of the seed can be remembered or forgotten depending on the initial inverse population. The excessive nonlinearity in the coherence amnesia regime can devastate the seed coherence, causing the oscillator to undergo a chaotic transition lasting hundreds of round trips before regaining coherence. Conversely, the oscillator converges in only a few round trips for the coherence memory regime. A heterodyne technique was developed to record the fast varying optical phase and characterize these two regimes. Dissipative soliton molecules were synthesized from external pulse pair seeds via the coherence memory pathway. In this case, a plateau of the generated pulse spacing independent from seed pulse spacing, i.e., amnesia of the seed spacing, was observed for close spaced seed pulse pairs. Moreover, we show that pulse seeds can be used for laser reconfiguration and pulse pattern control. 
Our work paves a way to control transient pulse dynamics and steady pulse forms on demand in mode-locked lasers.}

\begin{figure*}[ht]
\centering\includegraphics[width=\linewidth]{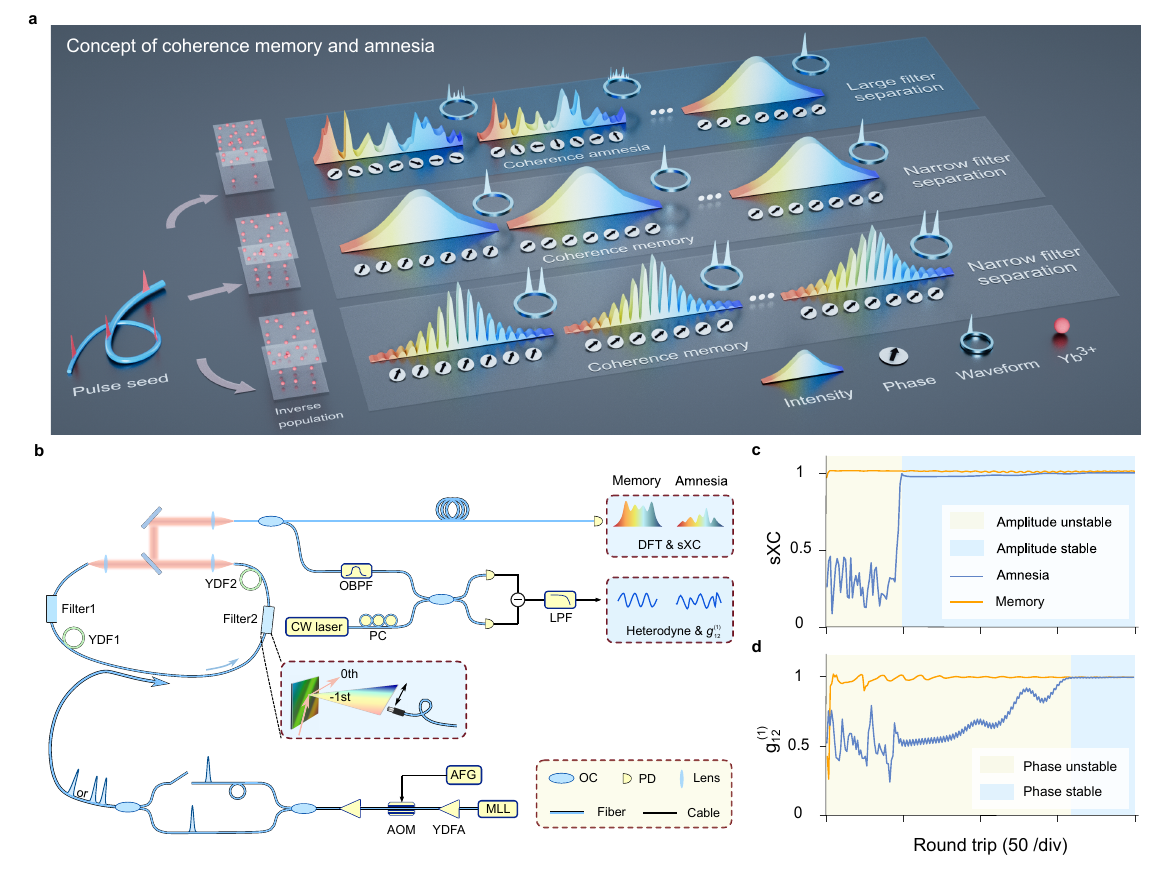}
\caption{{\bf Concept of coherence memory and amnesia in mode-locked oscillators.} {\bf a,} An illustration of the coherence transition pathways. By controlling the filter separation and pump power in the Mamyshev oscillator, the inverse population of the gain fiber can accumulate to different levels. Thus, the input pulse seed will be amplified to different levels. When the amplified power is very high, the excessive nonlinearity will distort the spectral amplitude and phase, resulting in `coherence amnesia'. When the inverse population is not very high, the seed coherence can be preserved and mode-locking can converge quickly, i.e., `coherence memory'. By this route, dissipative soliton molecules can be synthesized deterministically using a pulse-pair seed. {\bf b,} Experimental setup for the Mamyshev oscillator and the coherence transition pathway measurements. Both dispersive Fourier transform (DFT) and heterodyne beat are used to characterize the coherence transition dynamics. OC, optical coupler; AOM, acousto-optical modulator; MLL, mode-locked laser; AFG, arbitrary function generator; YDF(A), Yb-doped fiber (amplifier); OBPF, optical bandpass filter; PC, polarization controller. {\bf c,} Simulated power spectrum cross-correlation (sXC) and {\bf d,} complex spectral coherence $g_{12}^{(1)}$ evolution in the mode-locking buildup for the coherence memory and amnesia routes.}
\label{fig1}
\end{figure*}


\begin{figure*}[ht]
\centering\includegraphics[width=\linewidth]{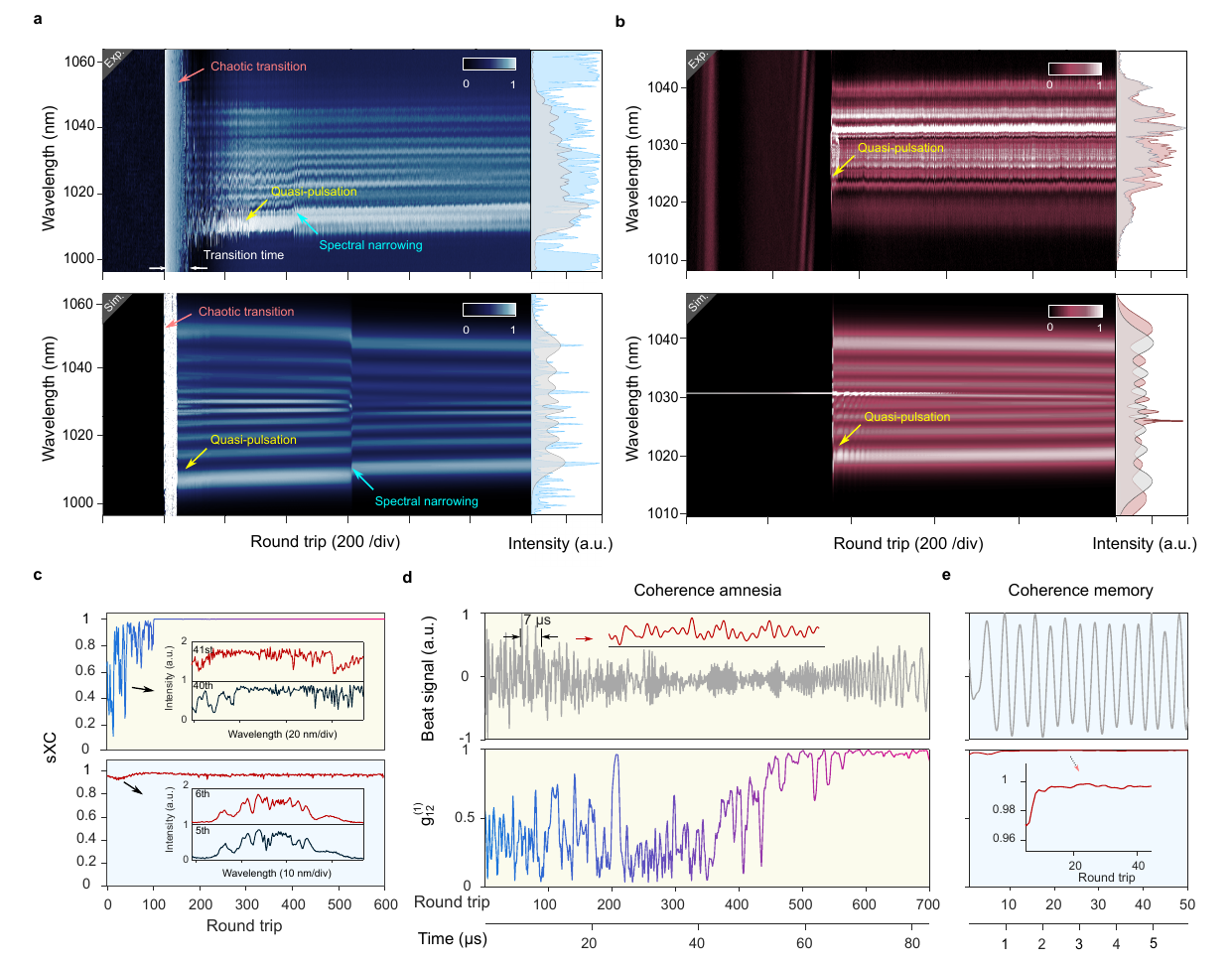}
\caption{{\bf Observation of coherence amnesia and memory in a Mamyshev oscillator.} {\bf a,} Experimentally measured single-shot spectra dynamics in the coherence amnesia regime and the corresponding simulation with a filter separation of 6 nm. The input seed undergoes a chaotic transition in the mode-locking buildup. The right panels show two examples of the chaotic (blue) and steady (gray) spectra. {\bf b,} Experimentally measured single-shot spectra dynamics in the coherence memory regime and the corresponding simulation with a filter separation of 3 nm. The right panel shows the spectra right after the pulse arrival (red) and in the steady state (gray). {\bf c,} Measured power spectrum cross-correlation (sXC) based on the single-shot spectra for the coherence amnesia (upper) and memory (bottom) regimes. The insets show examples of the single-shot spectra (the high power in the chaotic transition can saturate the photodetector). {\bf d, E,} Measured heterodyne beat signal with a CW laser for the coherence amnesia and memory regimes, respectively (the inset of panel {\bf d} is a zoom in of the irregular beat signal). The heterodyne signals are used to compute the spectral coherence $g_{1,2}^{(1)}$ shown in the bottom panels.}
\label{fig2}
\end{figure*}

\begin{figure*}[ht]
\centering\includegraphics[width=\linewidth]{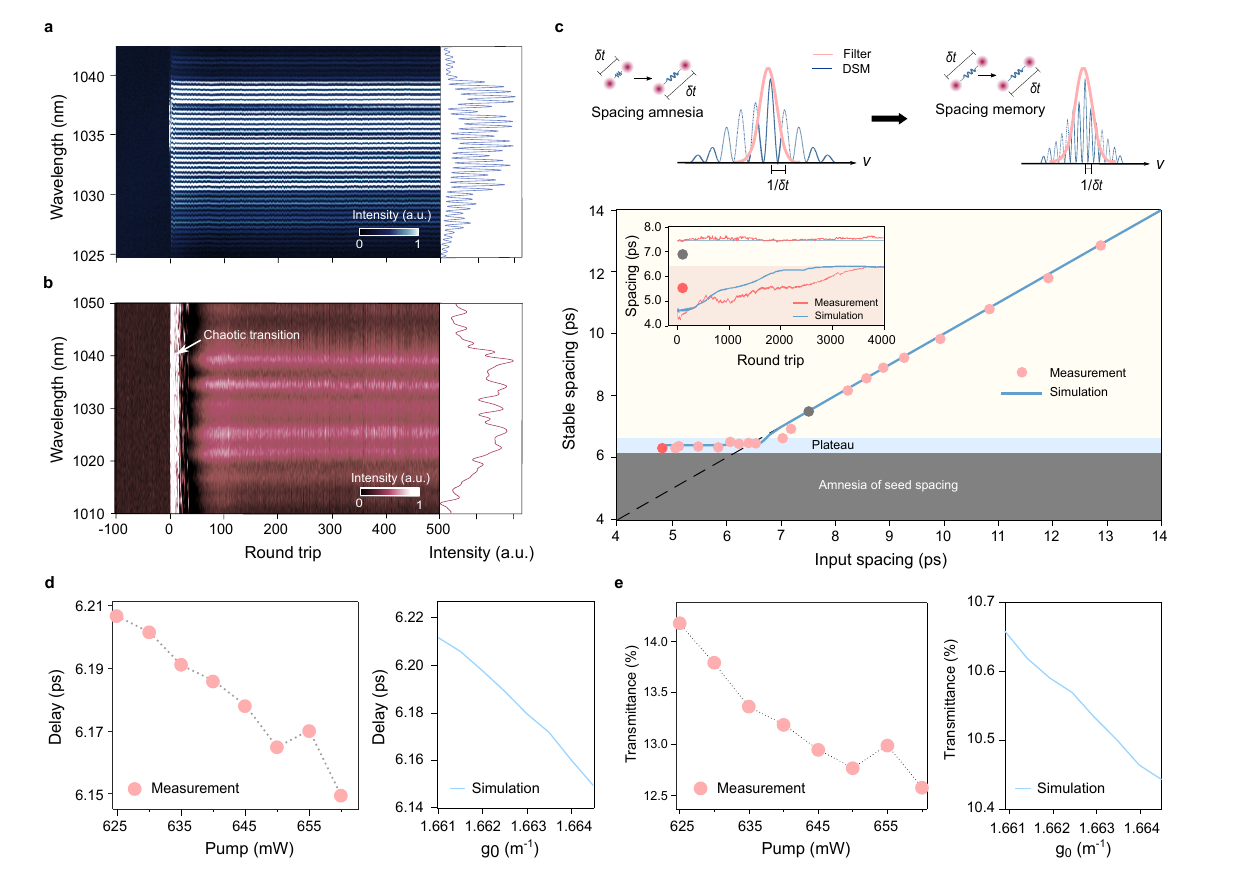}
\caption{{\bf Synthesize of dissipative soliton molecules (DSMs) with controlled pulse spacings.} {\bf a,} Experimentally measured single-shot spectra during the formation of a DSM seeded by a pulse pair spaced by 8.7 ps. The right panel shows the stabilized DSM power spectrum. {\bf b,} Single-shot spectra measured when the pulse pair seed undergoes the coherence amnesia pathway. The chaotic transition breaks the pulse pair and the laser stabilizes in a single-pulse state, with the spectrum shown in the right panel. {\bf c,} Measured stable DSM pulse spacing versus the seed pulse spacing. When the seed pulse spacing is relatively large, the spacing will be remembered. In contrast, a spacing plateau is measured and simulated for narrow spaced pulse seed. The inset shows the measured and simulated spacing change for different seed spacing cases (gray and red dots in the main panel). The upper panel is an illustration of the dependence of filter loss on pulse spacing. The pulse spacing only impacts the loss strongly, when the spacing is narrow and the spectral fringe is coarse.
{\bf d,} Measured and simulated pulse spacing of stable DSMs versus pump power or small signal gain $g_0$. {\bf e,} Measured and simulated filter transmission when tuning the pump power of the laser or $g_0$. }
\label{fig3}
\end{figure*}

\begin{figure*}[ht]
\centering\includegraphics[width=\linewidth]{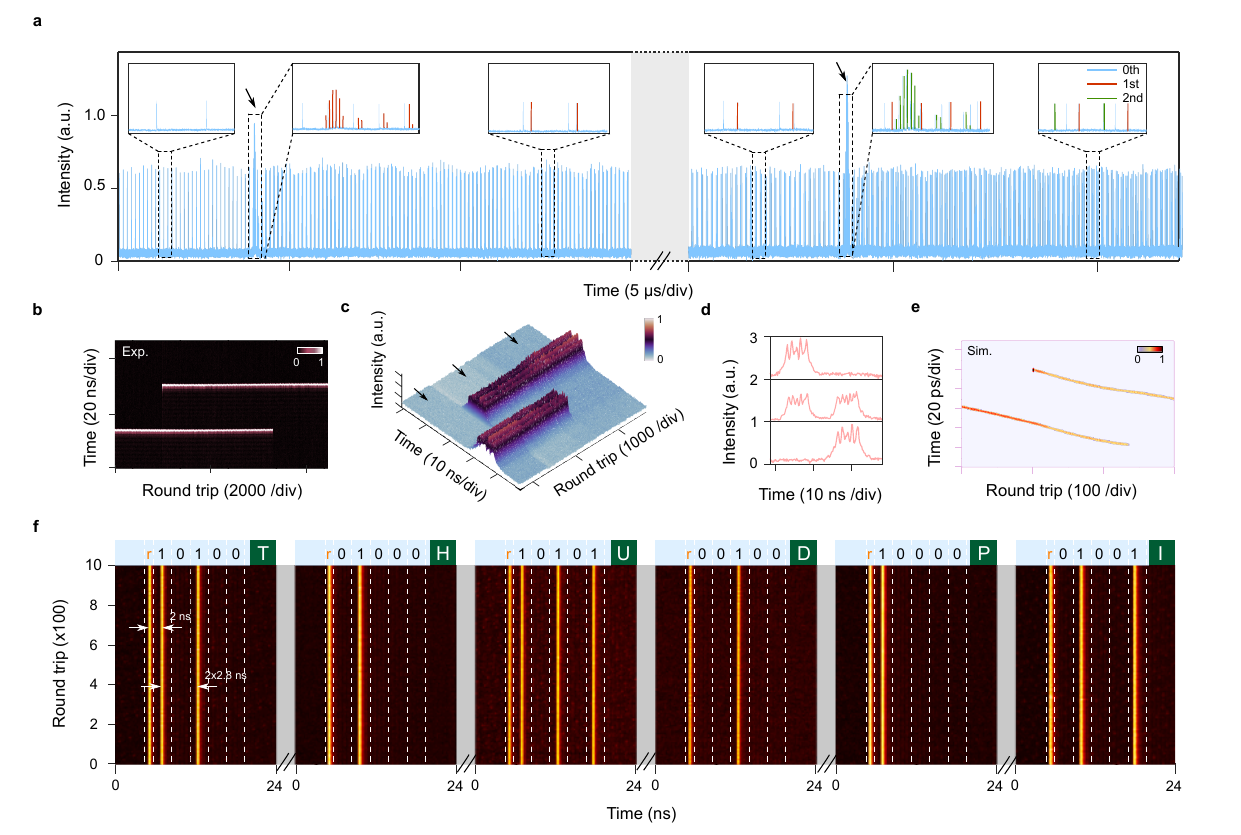}
\caption{{\bf Demonstration of pulse pattern reconfiguration and all optical bit storage.} {\bf a,} Temporal trace showing pulse writing into a laser already hosting a mode-locked pulse. The arrows show two moments when the new pulse seeds are injected. By injecting the pulse seeds, the laser can be reconfigured into a 3-pulse state. {\bf b,} Pulse erasing and transportation by an injected pulse seed without overlapping with the circulating pulse. {\bf c,} Single-shot spectrum dynamics corresponds to the temporal dynamics shown in panel {\bf b}. {\bf d,} Three example spectra at the round trips marked by arrows in panel {\bf c}. {\bf e,} Simulated pulse erasing by a new injected pulse. 
{\bf f,} Demonstration of pulse pattern control for optical bit storage. By preparing the pulse seed properly (see Supplementary Sec. 8), pulse streams representing a string `THUDPI' can be stored in the laser.}
\label{fig4}
\end{figure*}

\noindent{\bf Introduction.} Attaining coherence is a fundamental goal in optics \cite{schawlow1958infrared,Matei_PRL2017}. Passive mode-locking or self-organization among temporal modes is an indispensable approach to generate coherent dissipative soliton pulses \cite{Haus2000mode,grelu2012dissipative,Kippenberg_Science2018Review}. These mode-locked oscillators open the door for femtochemistry \cite{Zewail2000femtochemistry}, nonlinear spectroscopy \cite{cundiff2013optical,cheng2015vibrational}, exoplanet imager \cite{li2008laser,Hansch_Nature2012spectrograph}, and optical atomic clocks \cite{ludlow2015optical,Ye_Nature2022resolving}, greatly enhancing our ability to perceive the universe. Understanding the self-organization dynamics and the coherence transition pathways is essential to tailor the oscillators for better performance. On the other hand, optical oscillators provide a prism to look into the self-organization dynamics \cite{ilday2022universality}, which is a universal phenomenon occurring from macroscale sheep flocks \cite{gomez2022intermittent} to microscale multinucleon systems \cite{otsuka2019underlying}. An outstanding example in this effort is the real-time measurement of the mode-locking buildup dynamics from quantum noise in a titanium-sapphire femtosecond laser \cite{Jalali_NP2016resolving}. The experiment used the dispersive Fourier transform (DFT) technique \cite{goda2013dispersive} to measure non-repetitive spectrum dynamics on single-shot basis and showed how incoherent picosecond fluctuations go through abrupt spectral broadening and beating when transitioning into coherent femtosecond dissipative solitons \cite{Jalali_NP2016resolving}. The pioneering work is followed by intense studies on mode-locking buildup dynamics, which have established routes towards temporal modes organization from noise in mode-locked oscillators \cite{herink2017real,liu2018real,2018LPR_Zeng,Chen:18,liu2019revealing2,pu2020intelligent,pu2020intelligent,meng2021intracavity,zhao2021real,guo2021real,cao2022self}. 
However, controlling the transient mode-locking buildup dynamics and relative timing for stable multi-pulse states remains a significant challenge, since the birth of mode-locking. 

In this Article, we show that external pulse seeds provide a viable approach to address this challenge. Mode-locking in some oscillators hardly starts from quantum noise and trigger from external pulse seeds is needed. Fiber Mamyshev oscillators are such an example \cite{mamyshev1998all,liu2017megawatt}. The lasing threshold for them is above the mode-locking threshold \cite{liu2017megawatt}. Consequently, the initial inverse population will accumulate to different levels depending on the laser configuration before the arrival of the seed. Thus, it can be used a new knob to control the mode-locking buildup dynamics from coherent pulse seeds (Fig. \ref{fig1}a), which is distinct from the conventional quantum noise seeded cases.  
Experimentally, we observe a mode-locked oscillator can operate either in a coherence memory or a coherence amnesia regime with pulse seeds. Since the DFT technique cannot record the phase trajectory of a single pulse \cite{goda2013dispersive}, we also developed a heterodyne technique to record the coherence transition routes in real-time  (Fig. \ref{fig1}b). 
The observed coherence transition routes were used to control the multi-pulse forms in the oscillator. In a dissipative soliton molecule (DSM) synthesis experiment, we observed that relatively large spacings of the seeds can be remembered by the oscillator, while seeds with spacings below a threshold will always be attracted to a spacing plateau. Furthermore, we used  multi-pulse seeds with programmed spacings for a proof-of-concept optical buffer experiment. Numerical simulations are in qualitative agreement with the experiments and unveil the importance of the slow gain dynamics \cite{menyuk2007pulse} in underlying the coherence transition routes. Our work can stimulate further work on self-organization phenomena subject to ordered seeds. For instance, the observed coherence transition pathways can be readily examined for multi-mode fiber  Mamyshev oscillators \cite{Wise_OL2022multimode,cao2022self,cao2023spatiotemporal} and the vast traditional fiber lasers (see Supplementary Sec. 2). 

\noindent {\bf Coherence transition pathways.} The laser investigated is a Mamyshev oscillator with a repetition rate of 8.4 MHz (see Fig. \ref{fig1}b). It relies upon a pair of offset filters and nonlinear medium to promote pulse generation and forbid continuous wave (CW) emission \cite{liu2017megawatt}. The filter separation can be adjusted by tuning filter2 (a grating-based filter \cite{liu2017megawatt}, see Methods). Due to the offset filter pair, the laser operates below the threshold and emits spontaneous emission, when there is no external seed introduced. With the pulse seed injected (see Methods for seed control), it can be amplified and spectrally broadened in the fibers, and pass through the filter pairs. After circulating the cavity for some round trips, mode-locking can be established. 

There are two distinct routes towards mode-locking in the oscillator. When the initial inverse population is at a very high level, the injected pulse spectrum can be strongly broadened and become structured. Meanwhile, the spectral phase is randomized and the waveform becomes distorted in the time domain (see Fig. \ref{fig1}a). Both the amplitude and phase of the spectrum change randomly until reaching a stable mode-locking state. In other words, the phase coherence of the pulse seed is forgotten in the transition and we term it as the `coherence amnesia' regime. By contrast, if the initial inverse population is not very large, the spectrum will not be dramatically broadened and stable mode-locking can converge quickly. In this transition, phase coherence of the seed is well retained (see Fig. \ref{fig1}a). Thus, we term it as the `coherence memory' regime. DSMs can be deterministically generated via this route when seeding the oscillator with two close spaced pulses. 

To quantify the seed coherence transition routes, we first exploited the power spectrum cross-correlation (sXC) to show the variation of the power spectrum in the mode-locking buildup. sXC of the $N$th round trip is defined as the power spectrum cross-correlation with the $(N+1)$th round trip. The sXC for a simulated mode-locking buildup process is shown in Fig. \ref{fig1}c. In the coherence memory regime, the power spectrum is stabilized in only a few round trips, but it takes a much larger round trip number for the power spectrum to stabilize for the coherence amnesia regime. 

We further define the spectral coherence $g_{12}^{(1)}$ as
\cite{dudley2006supercontinuum},

\begin{equation}
 g_{12}^{(1)}(\lambda, N) =\frac{\left | \left \langle E_{1}^{*}(\lambda, N)E_{2}(\lambda, N+m) \right \rangle \right | }{\sqrt{\left \langle \left |E_{1}(\lambda, N) \right |^{2} \right \rangle\left \langle \left |E_{2}(\lambda, N+m) \right |^{2} \right \rangle } } ,
\end{equation}
where $E(\lambda, N)$ is the complex amplitude for wavelength $\lambda$ at the $N$th round trip, and angular brackets denote averaging over considered round trips for $m$=1,2,...,10. Thus, $g_{12}^{(1)}$ can describe how stable the spectral phase is between adjacent round trips. The simulated $g_{12}^{(1)}$ is plotted in Fig. \ref{fig1}d. It shows the phase coherence of the seed is retained for the coherence memory route, while the seed will go through a low coherence state lasting hundreds of round trips before regaining coherence. As an aside, comparison between Figs. \ref{fig1}c, D shows the phase takes a longer time to be synchronized than the stabilization of the amplitude. 

In experiments, we used the DFT and the heterodyne techniques to measure the spectral amplitude and phase transitions (Fig. \ref{fig1}b). By temporally stretching an ultrashort pulse using a long dispersive fiber, single-shot power spectra can be recorded by measuring the stretched pulses \cite{goda2013dispersive}, which facilitates the sXC measurement. By heterodyne beating the output with a CW laser, the transient complex field for a specific comb line can be measured, yielding the $g_{12}^{(1)}$ measurement (see Methods). 

\noindent {\bf Observation of coherence memory and amnesia.} The filter separation and pump power can be used to control the initial inverse population and coherence transition routes experimentally (see Supplementary Sec. 3 for a simulated phase diagram). The observed coherence memory and amnesia regimes are shown in Fig. \ref{fig2}. We had an acousto-optical modulator (AOM) to control the seed pulse train to generate a single pulsed seed event with a single-pulse energy of 60 pJ, and injected it into the Mamyshev oscillator via a 20\% coupler. When setting the filter separation at 6 nm and maintaining a pump power for YDF1 of 0.9 W, no CW lasing was observed prior to the seed injection. Thus, inverse population of the gain medium accumulated to a large number (see Fig. \ref{fig1}a). Upon the arrival of the seed pulse, it was amplified dramatically, resulting in a pronounced peak power. This elevated peak power led to dramatic broadening and distortion of the power spectrum (see the region labelled `chaotic transition' in Fig. \ref{fig2}a). 
Subsequently, the oscillator transitioned into a quasi-pulsation state, which is marked by quasi-periodic alterations in the power spectra \cite{cao2022observation}. After about 400 round trips, the spectrum underwent an abrupt narrowing and stabilized. 
The right panels of Fig. \ref{fig2}a show two examples of the spectra in the chaotic transition and the stabilized states. The observed dynamics including the chaotic transition, quasi-pulsation and spectral narrowing were all reasonably reproduced in simulations. Further analysis of the simulation shows the slow active fiber gain dynamics is critical to the observed coherence loss in the mode-locking buildup (see Supplementary Sec. 4). 



The coherence memory route was accessed by reducing the filter separation to 3 nm and the pump power to 0.7 W. The measured and simulated power spectrum dynamics is shown Fig. \ref{fig2}b. Since the filters overlap slightly, a weak relaxation oscillation was observed before the arrival of the seed pulse (see Fig. \ref{fig2}b) \cite{cao2022self}. 
As the CW lasing cannot be stretched in DFT, it occupies the whole measurement window, but this does not mean the spectrum is broad. Although the lasing is not continuous, it consumed the inverse population. When the seed arrived, it was only amplified mildly and the oscillator quickly converged to a coherent state without passing a chaotic transition. 
In this process, a weak quasi-pulsation was also observed; however, the spectral narrowing stage of the amnesia regime was absent, given that the spectrum did not undergo an extensive broadening stage (see also right panels of Fig. \ref{fig2}b), as observed in the amnesia state.

The single-shot spectra measured in Figs. \ref{fig2}a, B were used to compute the sXC in Fig. \ref{fig2}c. The power spectra alter on round trip basis during the chaotic transition for the coherence amnesia route (see the inset of Fig. \ref{fig2}c). Hence, the sXC is relatively low for the first 130 round trips, while the sXC is close to 1 upon the injection of the seed, which is consistent with the simulation in Fig. \ref{fig1}c. For the amnesia route, the detector can be saturated by the high output power in the chaotic transition, so as to measure the low power stable spectra. Thus, the measured sXC tends to be higher than the simulation. 

The heterodyne measurement results are shown in Figs. \ref{fig2}d, E. For the coherence amnesia regime, the heterodyne beat signal is irregular in the first hundreds of round trips (see the inset of Fig. \ref{fig2}d for a zoom-in of the signal), and regular oscillation was only observed after a relatively long transition. After retrieving the amplitude and phase of the heterodyne signal by the Hilbert transform, we obtained the complex field and $g_{\rm 12}^{(1)}$ for the corresponding comb line, which is plotted in the bottom panel of Fig. \ref{fig2}d. The spectral coherence remains relatively low in the first hundreds of round trips, and starts to ameliorate gradually after about 350 round trips. Relatively high coherence is only attained after about 500 round trips. Conversely, the heterodyne signal is regular right after the arrival of the seed pulse for the coherence memory regime (Fig. \ref{fig2}e). As a result, $g_{\rm 12}^{(1)}$ is close to 1 during the whole mode-locking buildup. The measured coherence transition in Figs. \ref{fig2}d, E agrees qualitatively with the simulation in Fig. \ref{fig1}d, and establishes two coherence transition pathways for the Mamyshev oscillator. The comparison with the measured sXC and $g_{\rm 12}^{(1)}$ verifies the spectral phase takes longer time to stabilize than the spectral amplitude.

\noindent {\bf Coherence memory for DSMs.} The observed coherence transition pathways can be leveraged to synthesize DSMs. The formation of DSMs in mode-locked lasers starting from noise exhibits rich and complex dynamics \cite{herink2017real,liu2018real}. Nevertheless, it remains challenging to control the DSM formation routes and the final pulse spacings in those lasers. 

Here, we constructed a fiber Mach-Zehnder interferometer (MZI) to prepare the seed pulse pair (Fig. \ref{fig1}b). 
DSMs can be generated via the coherence memory route with the filter separation and pump power set at 4.2 nm and 700 mW, respectively. Figure \ref{fig3}a shows an example of the measured single-shot spectrum dynamics in the formation of the DSM. No chaotic transition was observed, and the oscillator modes quickly organized to form a DSM with a spectrum exhibiting high contrast periodic fringes (right panel of Fig. \ref{fig3}a). Fourier transform of the power spectra (i.e., first order autocorrelation) shows that the spacing of the seed is remembered by the oscillator and the generated pulse spacing was stabilized at 8.7 ps right after the injection of the pulse pair seed (Fig. \ref{fig3}a). 
By contrast, DSMs cannot be synthesized even with the pulse pair seed, if they go through the coherence amnesia pathway with a filter separation of 5 nm and a pump power of 1.1 W (Fig. \ref{fig3}b). The chaotic transition will break the pulse pair seed into a chaotic temporal waveform, leading to a single pulse state. 


The spacing of the seed can be used to control the stabilized pulse spacing within the DSM. We plot the relationship between the measured final pulse spacing and the seed pulse spacing in Fig. \ref{fig3}c. When the spacing is larger than 6.2 ps, the stabilized pulse spacing stays the same with the seed pulse spacing. In other words, the spacing is remembered by the oscillator. The pulse spacing can also have an `amnesia' regime. When the seed pulse spacing was smaller than 6.2 ps, a spacing plateau at 6.2 ps was observed, independent of the seed spacing. The memory and amnesia of the pulse spacing were confirmed by simulations that are in excellent agreement with measurements (Fig. \ref{fig3}c). The inset of Fig. \ref{fig3}c shows examples of the DFT measured and simulated pulse separation change during the formation of the DSM starting from a seed spacing of 4.7 ps and 7.5 ps. For the pulse pair with a spacing of 4.7 ps, the pulse pair was gradually repulsed to the 6.2 ps plateau in thousands of round trips. Further analysis of the simulation shows that it is the peak power difference and the associated center frequency difference of the two pulses causing the repulsion (see Supplementary Sec. 5). Such a repulsion was not observed for the spacing memory state with a seeding spacing of 7.5 ps.

The existence of the spacing plateau can be understood as a requirement for gain-loss balance. The fringe period of a DSM power spectrum is 1/$\delta t$ ($\delta t$ is the pulse spacing, see top panel of Fig. \ref{fig3}c). Therefore, when $\delta t$ is relatively large, the fringe is dense and the transmission of the filter is not impacted significantly by $\delta t$. Thus, the gain-loss balance can be reached for various pulse spacings, and the seed pulse spacing can be remembered. When $\delta t$ is small, the fringes become coarse and the transmission of the filter becomes closely related to the pulse spacing. Therefore, seed spacing below a threshold is always pushed towards the plateau to have a similar filtering loss for gain-loss balance.   

To verify the gain-loss balance impact on the pulse spacing, we changed the pump power to measure $\delta t$ and filter2 transmission for the synthesized DSMs. When increasing the pump power, the pulse spacing plateau was observed to decrease from 6.21 ps to 6.15 ps, which also resulted in a decrease of the filter transmission. In the experiment, we collected the zeroth-order diffraction of the grating to measure the intracavity power and the transmission of the grating-based filter (see Fig. \ref{fig1}b and Supplementary Sec. 6), which is shown in Fig. \ref{fig3}e. The decreasing transmission (higher dissipation) with increasing pump power is consistent with the gain-loss balance picture. The measured pulse separation and transmission decrease agree well with the simulations in Figs. \ref{fig3}d, e. Our observation explicitly elucidates that the filter-dissipation is a critical mechanism shaping the DSM in the Mamyshev oscillator.

\noindent {\bf Pulse reconfiguration and bit storage.} Beyond DSMs, more complex pulse patterns are desired for many applications, including all optical bit storage \cite{leo2010temporal,pang2016all} and laser material ablation \cite{Ilday_Nature2016ablation}. 
However, access to pulse patterns with arbitrary separations in mode-locked lasers remains elusive, to our knowledge. External pulse seeds can give access to desired pulse patterns via the coherence memory route. 

In the first demonstration, we show that additional pulses can be written into the oscillator already accommodating mode-locked pulses. The laser initially operated in a single-pulse state with a filter separation of 5 nm. We used the AOM to switch on the pulse seed manually to write in new pulses. The first pulse seed was written at the moment indicated by the first arrow in Fig. \ref{fig4}a, and the laser converged to a 2-pulse state after about 4 round trips. Then, we introduced another seed event at the second arrow, and a 3-pulse state can be attained. The DFT measured single-shot spectra suggest that the seed pulses went through the coherence memory route (Supplementary Sec. 7), although the 6 nm filter separation tends to support the coherence amnesia route. This is because the initially orbiting pulse saturates the intracavity inverse population; thus, the new pulse did not experience the strong amplification and spectral broadening (Supplementary Sec. 7). 
Since we injected the new seed manually, the pulse separation was not controlled in the experiment. Precise electronic control of the seed event timing should enable spacing-defined pulse writing. 

The maximum pulse number that can be written into the oscillator depends on the filter separation and the pump power. Experiments show a maximum of 21-pulse can be possible when using a high pump power and a narrow filter separation (see Supplementary Sec. 7). When the oscillator only supports one pulse (with a filter separation of 5 nm and a pump power of 0.6 W), the external pulse seed can be used to erase the original pulse (see Fig. \ref{fig4}b). Since the pulse is not erased by cross-phase modulation (XPM) as in ref. \cite{pang2016all}, the control pulse does not need to overlap with the orbiting pulse. Effectively, the orbiting pulse is transported by about 20 ns by the external seed. The DFT measured spectra show the injected pulse went through the coherence memory route (Figs. \ref{fig4}c, d), and the final pulse has the same spectrum with the original one. Numerical simulations reproduce the erasing dynamics of the orbiting pulses by an external pulse (Fig. \ref{fig4}e). 


Finally, we wrote pulse sequences that may be used to buffer a string `THUDPI' into the Mamyshev oscillator (Fig. \ref{fig4}f). In the proof-of-concept demonstration, the order of these letters in the alphabet was encoded in the binary format. 6 pulse-slots were used for encoding a single letter, with the first slot used as a reference bit and the remaining 5 slots covering the 26 letters. The reference slot has an interval of 2 ns from the neighbour slot, while the others has an interval of 2.8 ns. By preparing the seed properly (see Supplementary Sec. 8), the encoded pulse sequences can be written successfully (Fig. \ref{fig4}f). We only wrote a letter into the laser in a single operation due to the complexity in preparing the pulse seed for a long string. Pulse shaping of the seeds in the Fourier domain \cite{Weiner_OC2011ultrafast} may be used to implement highly complex seed preparation. The experiment shows the potential of using Mamyshev oscillator for all optical bit storage and desired pulse pattern generation. 
The pulse timing control is on nanosecond scale in this experiment. For a higher storage capacity, more accurate pulse timing control will be required. The separation plateau observed in Fig. \ref{fig3}c and the round trip time will determine the ultimate storage depth of the oscillator.


\noindent {\bf Discussions.} We have observed two pathways towards mode-locking from pulse seeds, whose entrance can be controlled by the initial inverse population. By preserving the seed coherence, the temporal modes can quickly organize to form stable mode-locking pulses. Nevertheless, excessive nonlinearity can destroy the seed coherence, causing the oscillator to undergo a disordered stage before reorganizing into an ordered state. Besides controlling transient mode-locking buildup dynamics, the pulse seed can be used as a convenient tool to control the stable pulse timing, which is hard for pulses stemmed from quantum noise. The spacing within a DSM can be written into the oscillator by the seed spacing, as long as it is above a dissipation dynamics set threshold. Thus, the pulse pattern may be used to encode information and optical bit storage in the future. 
We anticipate that coherent seeds may also be used to control chip-based mode-locked oscillators e.g., quantum cascaded lasers \cite{Faist_NP2022dissipative} and quantum dot lasers \cite{Bowers_eLight2023prospects}. Our proposed quantities sXC and $g_{12}^{(1)}$ can also be used to characterize other photonic systems when transitioning from incoherent into coherent states. Beyond photonics, our work may inspire work on self-organization dynamics subject to organized excitation, instead of the commonly considered random actions \cite{gomez2022intermittent}. 

\vspace{3mm}

\noindent{\bf Materials and Methods}

\noindent{\bf Experiment setup.} The experiment is based on a Mamyshev oscillator centered at 1030 nm. Two arms of the oscillator comprise a 2.2 m (YDF1) and 1.7 m (YDF2) double-cladding Yb-fiber (LMA-10/125-YDF, Nufern) pumped by a 976 nm multimode pump, respectively. In the experiments, we fixed the pump for YDF2 at 1.1 W, and tuned the pump power for YDF1. All the remaining passive fibers in the cavity are HI1060 fibers. An isolator is used to ensure unidirectional propagation. To realize offset filtering, a bandpass dielectric filter centered at 1030 nm with 3 dB bandwidth of $\sim$3 nm is used. The other is formed by a 600 lines/mm reflective diffraction gratings (GR13-0610, Thorlabs) and a collimator \cite{liu2017megawatt,cao2022observation}. Thus, only a portion of the dispersed light can be collected, i.e., spectrally filtered. The passband of this filter can be adjusted by the grating orientation and the collimator position. The 0th-order diffraction light can be collected to monitor the intracavity pulse. A half waveplate is placed before the grating to maximize the diffraction efficiency. The filter bandwidth is measured to be $\sim$2 nm. 

The pulses are sampled by a polarization beam splitter (PBS) as the output. The pulse train is detected by a 5 GHz photodetector (PD1) and a 12 GHz photodetector (PD2), and digitized by a real-time oscilloscope (Keysight DSOS804A) with a bandwidth of 8 GHz. The averaged optical spectra are recorded by an optical spectrum analyzer (OSA, Agilent 86142B) with a resolution of 0.06 nm. The DFT path includes a 12.2 km SMF, corresponding to a net group delay dispersion of $\sim$360 ps/nm around 1030 nm. The resolution of single-shot DFT spectra is estimated to be 0.29 nm \cite{goda2013dispersive}. For the heterodyne measurement, the laser output is filtered by a optical bandpass filter centered at 1030 nm with a 3 dB bandwidth of 3 nm, and combined with a CW laser at 1030 nm by a coupler. The output is received by a balanced photodetector (Thorlabs PDB 450C) and low-pass filtered (\textless 4 MHz).

The seeds are from a home-built all normal dispersion fiber laser \cite{chong2006all} which delivers a chirped pulse with 4 ps duration and 64 MHz repetition rate. The seeds are amplified by an ytterbium doped fiber amplifier before entering an AOM driven by an arbitrary waveform generator. The 1st diffraction of the AOM is sent into a free space MZI to generate a single pulse (blocking an arm of the MZI) or a two-pulse seed. A translation stage is inserted to one of the arms to control the spacing of the two-pulse seed with a resolution of $\sim$0.7 fs.

\vspace{1mm}
\noindent{\bf Simulation methods.} The simulation is based on the lumped model, with the pulse passing the cavity components shown in Fig. \ref{fig1}b in sequence. Simulations start from a pulsed condition, with a single-pulse input for Figs. \ref{fig1}, \ref{fig2}, and a two
-pulse input for Fig. \ref{fig3}. Pulse propagation within the optical fibers is modeled by combining the generalized nonlinear Schrodinger equation and a phenomenological gain rate equation \cite{menyuk2007pulse}, that can be written as,
\begin{equation}
\setlength{\belowdisplayskip}{0.25pt}
  \frac{\partial A}{\partial z}+i\frac{\beta_2}{2}\frac{\partial^2A}{\partial t^2}-\frac{\beta_3}{6} \frac{\partial^3A}{\partial t^3}=\frac{g}{2}A+i\gamma|A|^2A+\frac{g}{2\Omega^2_g}\frac{\partial^2A}{\partial t^2},
\label{nlse}
\end{equation}
\begin{equation}
\setlength{\abovedisplayskip}{0.25pt}
 \frac{dg}{dT}=\frac{g_0-g}{\tau_f}-\frac{1}{\tau_f}\frac{gW}{W_{\rm sat}},
\label{gain}
\end{equation}
where $A$ is the electric field envelope of the pulse; $\beta_2$ and $\beta_3$ are the second-order and third-order dispersion, respectively; $\gamma$ is the Kerr nonlinearity coefficient; $t$ is the fast time and $z$ is the propagation distance; $\Omega_g$ is the gain bandwidth and $g$ is the saturated gain, whose motion is governed by Eq. \ref{gain}. $g_0$ and $W_{\rm sat}$ in Eq. \ref{gain} are the small signal gain and the gain saturation energy, respectively, and $W$ is the total intracavity pulse energy; $T=z/v_g$ is the slow time ($v_g$ is the group velocity) and $\tau_{f}$ is the fluorescence lifetime of the gain medium. The  used $\tau_{f}$ can be shorter than the actual lifetime; therefore, the simulated dynamics tends to be faster than measurement. For passive fibers, small signal gain is chosen as $g_0$=0 m$^{-1}$. After pulse propagation in one of the arm, it is filtered and enters the other arm, and passes through the other filter, closing the loop for one round trip. The simulation iterates till stabilization of the pulse. Further details of the simulation model and used parameters are listed in Supplementary Sec. 1.


\vspace{1 mm}
\noindent \textbf{Data Availability}
The data that supports the plots within this paper and other findings is available upon reasonable request from the authors.

\vspace{1 mm}
\noindent \textbf{Acknowledgments}
We acknowledge discussions with Fan O. Wu and Logan Wright. This work is supported by the National Natural Science Foundation of China (61975090, 62175127, 62250071, 62375150), the Tsinghua University Initiative Scientific Research Program (20221080069), and the Tsinghua-Toyota Joint Research Fund.

\vspace{1 mm}
\noindent\textbf{Author Contributions} All the authors conceived the project. B.C. led the experiment with assistance from Z.L. and C.G., B.C. and Z.L. conducted the simulation. The project is supervised by C.B. 
\vspace{1 mm}

\noindent \textbf{Competing Interests} The authors declare no competing interests.



\bibliography{apssamp}

\end{document}